\documentstyle[graphicx,aps,multicol]{revtex}
\draft

\begin{document}

\title{Gamma-ray strength function and pygmy resonance in rare earth nuclei}
\author{A.~Voinov\footnote{email: voinov@nf.jinr.ru}}
\address{Frank Laboratory of Neutron Physics, Joint Institute of Nuclear
Research, 141980 Dubna, Moscow reg., Russia}
\author{M.~Guttormsen, E.~Melby, J.~Rekstad, A.~Schiller\footnote{Corresponding
author, email: Andreas.Schiller@fys.uio.no} and S.~Siem}
\address{Department of Physics, University of Oslo, N-0316 Oslo, Norway}
\maketitle

\begin{abstract}
The $\gamma$-ray strength function for $\gamma$ energies in the 1--7~MeV region
has been measured for $^{161,162}$Dy and $^{171,172}$Yb using the 
($^3$He,$\alpha\gamma$) reaction. Various models are tested against the
observed $\gamma$-ray strength functions. The best description is based on the
Kadmenski{\u{\i}}, Markushev and Furman E1 model with constant temperature and
the Lorentzian M1 model. A $\gamma$-ray bump observed at $E_\gamma\sim 3$~MeV 
is interpreted as the so-called pygmy resonance, which has also been observed 
previously in (n,$\gamma)$ experiments. The parameters for this resonance have 
been determined and compared to the available systematics.
\end{abstract}


\begin{multicols}{2}

\section{Introduction}

The concept of $\gamma$-ray strength functions was introduced in the 
fundamental work of Blatt and Weisskopf \cite{Blatt}. They showed that the 
square of the $\gamma$-transition matrix element connecting highly excited
states is proportional to the level spacing $D_i$ of the initial states $i$ 
with equal spin and parity. Therefore, the ratio of the partial radiative width
$\Gamma_i$ of the states $i$ (which is connected to $\gamma$ transitions with 
transition energy $E_\gamma$ and populating some low lying levels) and $D_i$ 
was suggested to be important for the description of $\gamma$ transitions in 
the continuum. The corresponding model-independent definition of the 
$\gamma$-ray strength function\footnote{In literature, also called the 
radiative strength function.} is given by 
$f_{{\mathrm{X}}L}=\Gamma_i/(E_{\gamma}^{2L+1}D_i)$, where $L$ is the 
multipolarity of the $\gamma$ transition and X refers to the electric or 
magnetic character of the $\gamma$ transition. The $\gamma$-ray strength 
function is now considered as a measure for the average electromagnetic 
properties of nuclei and is fundamental for the understanding of nuclear 
structure and reactions involving $\gamma$ rays.

Experimentally, the main information on the $\gamma$-ray strength function has
been obtained from the study of photoabsorption cross-sections \cite{DB88}. It
is commonly adopted that the E1 strength function is determined by the
properties of the giant electric dipole resonance (GEDR) around its resonance
energy, typically $E_\gamma\sim$ 10--20~MeV\@. However, serious lack of
information persists at lower $\gamma$-ray energies. It was assumed that the
tail of the Lorentzian describing the GEDR determines the E1 strength function
at these energies. The only experimental data on the E1 $\gamma$-ray strength
function between compound states with $\gamma$-ray energies below 2~MeV have
been obtained using the $^{143}$Nd(n,$\gamma\alpha)$ reaction \cite{Popov}.
These data show that the extrapolation of the GEDR to low energies fails to
describe the experimental values of the E1 strength function and indicates a
finite value of $f_{{\mathrm{E}}1}$ in the limit $E_\gamma\rightarrow 0$. As a
result, a model for the E1 strength function was developed by 
Kadmenski{\u{\i}}, Markushev and Furman (KMF) \cite{Kadm1} which takes into
account the energy and temperature dependence of the GEDR width. Today, this
model and its empirical modifications \cite{Kope2} are frequently used in the
description of experimental data, but at the same time, it needs additional
experimental verification.

The E1 strength function is not solely governing the $\gamma$-ray emission for
lower $\gamma$-ray energies. Other multipolarities, and especially the M1
strength function, play important roles as well. The experimental information
on the $\gamma$-ray strength of M1 transitions is more scarce. It is commonly
assumed that the M1 strength is well described by the Weisskopf model 
\cite{Blatt}, where the dipole $\gamma$-ray strength function is independent of
the $\gamma$ energy. But some experiments indicate the existence of an M1 giant
resonance originating from spin-flip excitations in the nucleus \cite{Bohr}.
Also, the analysis of $\gamma$-ray spectra from (n,$\gamma)$ reactions
\cite{Ko93} indicates that the use of the M1 giant dipole resonance model gives
a better fit to the experimental data than the Weisskopf model.

Special attention \cite{Igas1,Berg1,Bart1,Barr1,Joly1} has been devoted to the
anomalous bump found in the $\gamma$-ray spectra of the (n,$\gamma)$ and 
(d,p$\gamma)$ reactions at low energies. The same bump has probably also been 
observed in the ($^3$He,$\alpha\gamma$) reaction \cite{Gutt2}. A previous work 
\cite{Igas1} shows that the energy of the $\gamma$-ray bump increases with 
neutron number in the $N=$~82--126 region. The bump is called the pygmy
resonance due to the considerably lower strength compared to the GEDR\@. The
pygmy resonance has first been explained by the enhancement of the E1 strength
function \cite{Igas1}. However, one can not rule out a possible M1 character
connected to orbital M1 strength (scissors mode) in nuclei, which was first
observed in electron scattering experiments \cite{Riht}.

The $\gamma$-ray strength function is difficult to measure in the 
$\gamma$-decay between highly excited states, since the decay rate also depends
on the number of accessible levels. The analysis of $\gamma$-ray strength 
functions from the spectra of (n,$\gamma)$ reactions \cite{Ko93} shows that the
results depend crucially on the level density model employed. Therefore, 
conclusions based on a certain level density formula should be considered as 
preliminary and thus, need confirmation.

Recently \cite{Hend1,Tvet1,Schi1}, a new experimental technique has been
developed, based on a set of primary $\gamma$-ray spectra measured at
consecutive excitation energies $E$ in light-ion reactions with one charged
ejectile. The technique allows to disentangle the $\gamma$-ray spectra into a
$\gamma$-energy dependent function $F(E_{\gamma})$ (which, as will be shown 
below, can be uniquely connected to the $\gamma$-ray strength function) and
the level density $\rho(E)$. It makes it possible to study the $\gamma$-ray
strength function and level density independently from each other, in contrast 
to what can be done by using radiative neutron capture techniques. In previous 
works \cite{Melb1,Schi2,Gutt1} the extracted level densities $\rho$ have been 
utilized to deduce thermodynamical properties for several rare earth nuclei. In
this paper, however, we will focus on the $F(E_{\gamma})$ function.

The experimental method is described in Sect.\ II\@. In Sect.\ III we give a
short outline of the models used to describe the experimental data, and in
Sect.\ IV we compare various predictions to the experimental data. Our results
are also compared to data from the (n,$\gamma)$ reaction performed by others.
Conclusions are given in Sect.\ V.

\section{Experimental method}

The experiments were carried out with 45~MeV $^3$He-projectiles at the Oslo
Cyclotron Laboratory (OCL). The particle-$\gamma$ coincidences are measured
with the CACTUS multidetector array \cite{Gutt3} using the 
($^3$He,$\alpha\gamma$) reaction on $^{162,163}$Dy and $^{172,173}$Yb
self-supporting targets. The charged ejectiles were detected with eight
particle telescopes placed at an angle of 45$^{\circ}$ relative to the beam
direction. An array of 28~NaI $\gamma$-ray detectors with a total efficiency of
$\sim$15\% of $4\pi$ surrounded the target and particle detectors.

The experimental extraction procedure and the assumptions made are described in
Refs.\ \cite{Hend1,Schi1} and references therein. From the 
$\alpha$-$\gamma$-coincidences, spectra of the total $\gamma$-ray cascade can
be sorted out according to the initial excitation energy $E$. These spectra are
the basis for making the first generation (or primary) $\gamma$-ray matrix
$P(E,E_{\gamma})$, which is factorized according to the Brink-Axel hypothesis
\cite{Brin1,Axel1}
\begin{equation}
P(E,E_{\gamma})\propto F(E_{\gamma})\rho(E-E_{\gamma}).
\label{eq:matrix}
\end{equation}
Here, $F$ and $\rho$ are the $\gamma$-ray energy dependent factor and the level
density, respectively. It is now possible to determine $F$ and $\rho$ by an
iterative procedure. The first trial function for $\rho$ is simply taken as a
straight line and the corresponding $F$ is determined by Eq.~(\ref{eq:matrix}).
Then, a $\chi^2$ minimum is calculated for each data point of $F$ and $\rho$,
keeping the others fixed. This procedure is repeated about 50~times, until a
global least square fit to the $\sim$1400 data points of the $P(E,E_{\gamma})$
matrix is achieved.

It has been shown \cite{Schi1} that if one solution for $F$ and $\rho$ has been
found, functions of the form
\begin{eqnarray}
\tilde{\rho}(E-E_\gamma)&=&A\,\rho(E-E_\gamma)\,\exp(\alpha\,[E-E_\gamma])\\
\tilde{F}(E_\gamma)&=&B\,F(E_\gamma)\,\exp(\alpha\,E_\gamma)
\label{eq:solut}
\end{eqnarray}
give exactly the same fit to the $P(E,E_{\gamma})$ matrix. The values of $A$,
$B$ and $\alpha$ can be determined by additional conditions. The $A$ and
$\alpha$ parameters are used for absolute normalization of the level density
$\rho$: They are adjusted to reproduce ({\em i}) the number of levels observed
in the vicinity of the ground state and ({\em ii}) the neutron resonance
spacing at the neutron binding energy $B_n$. Further details on the extraction
procedure and the simulation of errors are given in Ref.\ \cite{Schi1}. In the
following we will concentrate on the $\gamma$-ray energy dependent function
$F(E_{\gamma})$ and its normalization.

We assume that the main contributions to the derived $F$ function are from E1
and M1 $\gamma$-transitions and that the accessible levels of positive and 
negative parity are equal in number for any energy and spin i.e.
\begin{equation}
\rho(E-E_\gamma,I_f,\pm\Pi_f)=\frac{1}{2}\rho(E-E_\gamma,I_f).
\end{equation}
Thus, the observed $F$ is expressed by a sum of the E1 and M1 $\gamma$-ray 
strength functions only
\begin{equation}
BF(E_{\gamma})=[f_{\mathrm{E1}}(E_{\gamma})+f_{\mathrm{M1}}(E_{\gamma})]
E_{\gamma}^3,
\label{eq:prim1}
\end{equation}
where $B$ is the unknown normalization constant. Our experiment does not 
provide the possibility to derive the absolute normalization of $F(E_{\gamma})$
(see Eq.\ (\ref{eq:solut})), therefore, the normalization constant has to be
determined from other experimental data. The experimental, average total 
radiative width of neutron resonances $\langle\Gamma_\gamma\rangle$ at the 
neutron binding energy $B_n$ can e.g.\ be written in terms of $F$. To show 
this, we start with Eq.\ (3.1) of Ref.\ \cite{Kope2}
\begin{eqnarray}
\langle\Gamma_\gamma(E,I,\Pi)\rangle&=&\frac{1}{\rho(E,I,\Pi)}
\sum_{{\mathrm{X}}L}\sum_{I_f,\Pi_f}\int_{E_\gamma=0}^E{\mathrm{d}}E_\gamma
\nonumber\\
&&E_\gamma^{2L+1}f_{{\mathrm{X}}L}(E_\gamma)\rho(E-E_\gamma,I_f,\Pi_f)
\label{eq:ggdef}
\end{eqnarray}
where $\langle\Gamma_\gamma(E,I,\Pi)\rangle$ is the average total radiative 
width of levels with energy $E$, spin $I$ and parity $\Pi$. The summations and
integration are going over all final levels with spin $I_f$ and parity $\Pi_f$
which are accessible by $\gamma$ radiation with energy $E_\gamma$, 
multipolarity $L$ and electromagnetic character X\@. If we, again, assume that
only dipole radiation contributes significantly to the sum and that the number 
of accessible levels with positive and negative parity are equal, we obtain, by
combining Eqs.\ (\ref{eq:prim1}) and (\ref{eq:ggdef}), the average total 
radiative width of neutron s-wave capture resonances with spins $I_t\pm1/2$ 
expressed in terms of the $F(E_\gamma)$ function
\begin{eqnarray}
\lefteqn{\langle\Gamma_\gamma(B_n,I_t\pm1/2,\Pi_t)\rangle=\frac{1}
{2\rho(B_n,I_t\pm1/2,\Pi_t)}\int_{E_\gamma=0}^{B_n}{\mathrm{d}}E_\gamma}
\nonumber\\
&&BF(E_\gamma)\rho(B_n-E_\gamma)\sum_{J=-1}^1g(B_n-E_\gamma,I_t\pm1/2+J),
\label{eq:norm}
\end{eqnarray}
where $I_t$ and $\Pi_t$ are the spin and parity of the target nucleus in the
(n,$\gamma$) reaction and $\rho$ is the experimental level density. 
Furthermore, we have expressed $\rho$ as the product of the total level 
density, summed over all spins and the spin distribution $g$. The spin 
distribution of the level density is given by \cite{GC65}
\begin{equation}
g(E,I)=\frac{2I+1}{2\sigma^2}\exp\left[-(I+1/2)^2/2\sigma^2\right],
\end{equation}
where $\sigma$ is the excitation-energy dependent spin cut-off parameter. The 
spin distribution is normalized to $\sum_{I}g\approx 1$. The experimental value
of the average total radiative width of neutron resonances 
$\langle\Gamma_\gamma\rangle$ is then the weighted sum of contributions with 
$I_t\pm1/2$ according to Eq.\ (\ref{eq:norm}).

Because of methodical difficulties, the functions $F(E_{\gamma})$ and $\rho(E)$
can not be determined experimentally in the interval $E_\gamma<1$~MeV and
$E>B_n-1$~MeV, respectively. In addition, the data at the highest $\gamma$-ray
energies, $E_{\gamma}>B_n-1$~MeV, suffer from poor statistics. Therefore,
extrapolations of $F$ and $\rho$ were necessary in order to calculate the
integral in Eq.\ (\ref{eq:norm}). The contribution from the extrapolation to 
the total radiative width in Eq.~(\ref{eq:norm}) does not exceed 15\%, thus the
errors due to a possibly poor extrapolation are expected to be of minor 
importance.

\section{Models for E1 and M1 radiation}

There have been developed several models for the $\gamma$-ray strength
functions $f_{{\mathrm{X}}L}$. The theories behind the models are complicated,
and will not be outlined here. However, the resulting strength functions can be
written in simple analytical forms. In this work, we have tested various E1 and
M1 models. For E1 $\gamma$-transitions these are:
\begin{itemize}
\item The standard giant electric dipole resonance (GEDR) model based on the
Brink-Axel approach \cite{Brin1,Axel1}
\begin{equation}
f_{\mathrm{E1}}(E_\gamma)=\frac{1}{3\pi^2\hbar^2c^2}\frac{\sigma_{\mathrm{E1}}
E_\gamma\Gamma_{\mathrm{E1}}^2}{(E_\gamma^2-E_{\mathrm{E1}}^2)^2+E_\gamma^2
\Gamma_{\mathrm{E1}}^2}
\label{eq:lorentz}
\end{equation}
where $\sigma_{\mathrm{E1}}$, $\Gamma_{\mathrm{E1}}$ and $E_{\mathrm{E1}}$ are
the giant electric dipole resonance parameters derived from photoabsorption
experiments.\footnote{The constant $1/(3\pi^2\hbar^2c^2)$ equals 
$8.6\times10^{-8}$~mb$^{-1}$MeV$^{-2}$.}
\item The model of Kadmenski{\u{\i}}, Markushev and Furman (KMF) \cite{Kadm1}
\begin{equation}
f_{\mathrm{E1}}(E_\gamma)=\frac{1}{3\pi^2\hbar^2c^2}\frac{0.7
\sigma_{\mathrm{E1}}\Gamma_{\mathrm{E1}}^2(E_\gamma^2+4\pi^2T^2)}
{E_{\mathrm{E1}}(E_\gamma^2-E_{\mathrm{E1}}^2)^2},
\label{eq:Kad}
\end{equation}
where $T$ is the temperature of the nucleus which is usually determined as
$T=\sqrt{U/a}$ with $U$ being the shifted excitation energy and $a$ the level
density parameter. The energy and temperature dependent width of the GEDR in
this model is expressed by
\begin{equation}
\Gamma_{\mathrm{E1}}(E_\gamma,T)=\frac{\Gamma_{\mathrm{E1}}}{E_{\mathrm{E1}}^2}
(E_\gamma^2+4\pi^2T^2).
\label{eq:temper}
\end{equation}
These expressions are developed in the framework of a collisional damping model
for $E_\gamma<E_{{\mathrm{E}}1}$ and although they should hold for 
$T\ll 2$~MeV, the absence of thermal shape fluctuations in the model limits 
their validity to $T<1$~MeV\@.
\end{itemize}
For deformed nuclei, the giant dipole resonance is split into two components,
hence the sum of two strength functions with different GEDR parameters has been
employed.

For M1 $\gamma$-transitions we are testing:
\begin{itemize}
\item The adjusted single-particle model of Weisskopf \cite{Blatt} where 
$f_{\mathrm{M1}}(E_\gamma)$ is independent of $E_\gamma$ and the absolute value
of $f_{\mathrm{M1}}$ has been taken from $f_{\mathrm{M1}}/f_{\mathrm{E1}}$ 
systematics close to the neutron binding energy \cite{Kope3}.
\item A Lorentzian based on the existence of a giant magnetic dipole resonance
(GMDR) which is assumed to be related to the spin-flip transition between
single-particle states \cite{Bohr}. The $\gamma$-ray strength function in this
case is determined by
\begin{equation}
f_{\mathrm{M1}}(E_\gamma)=\frac{1}{3\pi^2\hbar^2c^2}\frac{\sigma_{\mathrm{M1}}
E_\gamma\Gamma_{\mathrm{M1}}^2}{(E_\gamma^2-E_{\mathrm{M1}}^2)^2+E_\gamma^2
\Gamma_{\mathrm{M1}}^2}.
\label{eq:M1}
\end{equation}
\end{itemize}
In the following, we will compare these models to the experimental findings.

\section{Results and discussion}

\begin{figure}\centering
\includegraphics[totalheight=17.9cm]{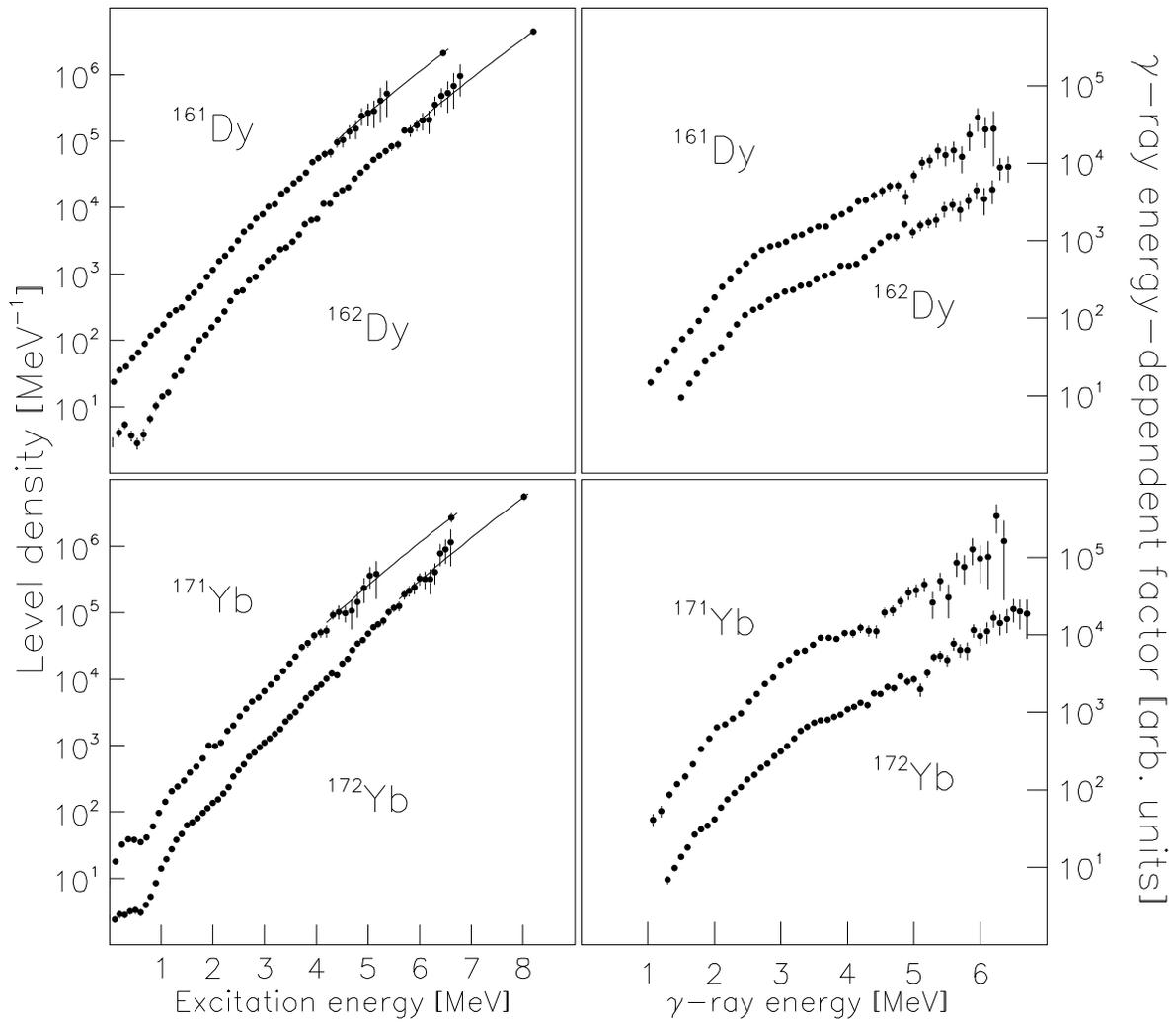}
\caption{The observed level density $\rho$ and the $\gamma$-ray energy
dependent factor $F$ for $^{161,162}$Dy and $^{171,172}$Yb.}
\label{fig:level}
\end{figure}

Figure \ref{fig:level} shows the experimentally extracted $\gamma$-ray energy
dependent factor $F$ and the level density $\rho$ for $^{161,162}$Dy and
$^{171,172}$Yb. These data are the very same as in Ref.\ \cite{Schi2}, except
that the data of $^{171}$Yb have been retuned by adjusting the parameters $A$
and $\alpha$ to fit the level density based on known discrete levels at low
excitation energy. We recognize that the shape of the unnormalized $F$
functions are rather equal for neighboring isotopes indicating that $F$ is a
slowly varying function of mass number.

In the extraction procedure, we have used $\gamma$-ray spectra from excitation
energy bins between 4 and 8~MeV\@. This span in excitation energy corresponds 
to a temperature region of 0.5 to 0.7~MeV for the initial states, and 0.4 to
0.6~MeV for the final states. The spin window in the pick-up reaction is
$2-6\,\hbar$, which is assumed to be approximately equal for initial and final
states. Hence, the discussion below concerns average properties of nuclei for
$T\sim 0.5$~MeV and $I\sim 4\,\hbar$ with the assumption of equal density of
positive and negative parity states.

\begin{figure}\centering
\includegraphics[totalheight=17.9cm]{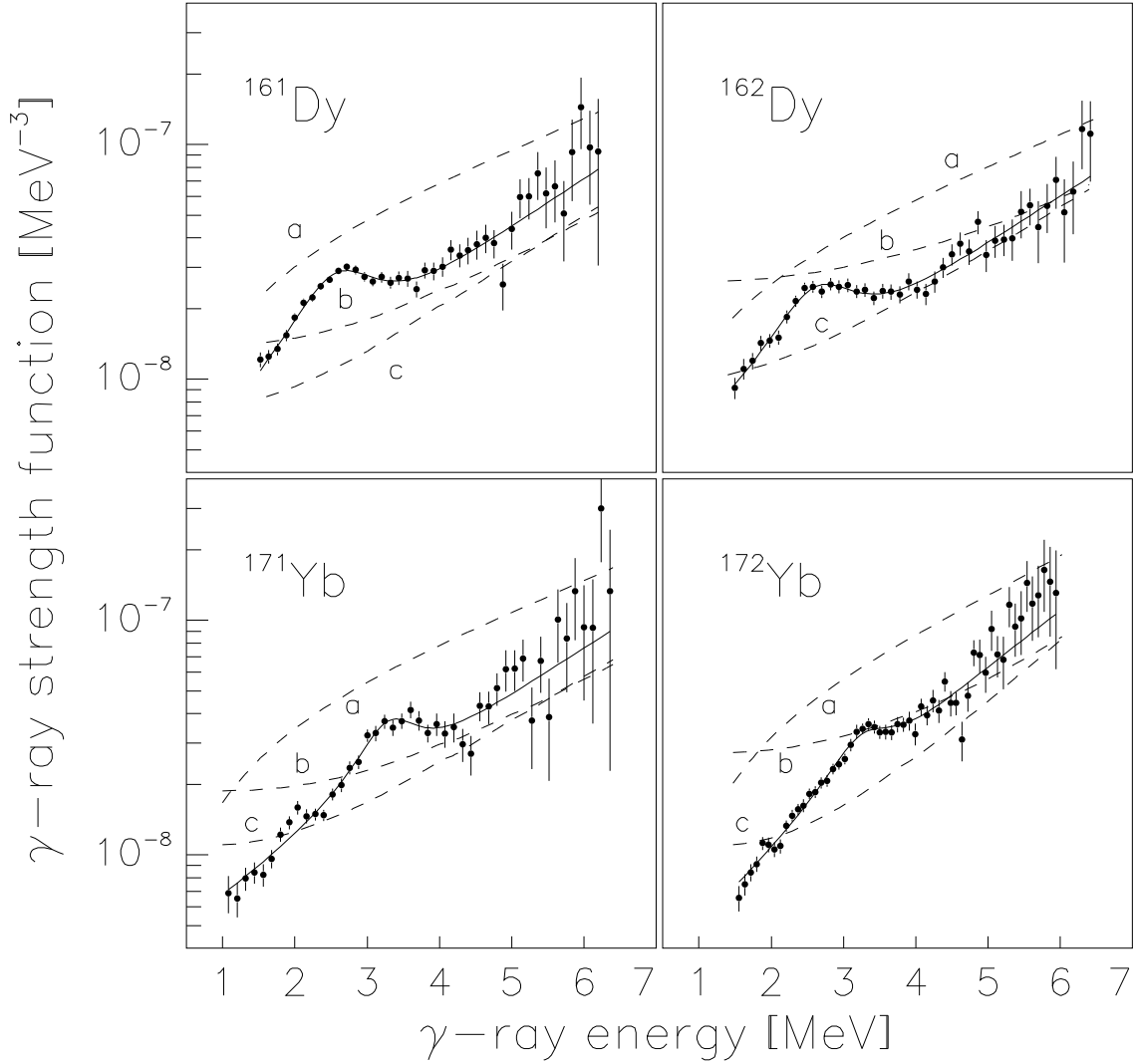}
\caption{The observed $\gamma$-ray strength functions (data points with error
bars) for $^{161,162}$Dy and $^{171,172}$Yb. The dashed curves are calculations
where a) denotes the Lorentzian GEDR model [Eq.\
({\protect{\ref{eq:lorentz}}})], b) the KMF model [Eq.\
({\protect{\ref{eq:Kad}}})] plus a Weisskopf estimate for M1 transitions, and
c) the KMF model [Eq.\ ({\protect{\ref{eq:Kad}}})] plus a Lorentzian GMDR model
[Eq.\ ({\protect{\ref{eq:M1}}})]. For b) and c), the temperature is given by
$T=\sqrt{U/a}$. The solid curves are the KMF model with constant temperature
[Eq.\ ({\protect{\ref{eq:Kad}}})] plus a Lorentzian GMDR model [Eq.\ 
({\protect{\ref{eq:M1}}})] plus a Lorentzian pygmy resonance model [Eq.\ 
({\protect{\ref{eq:py}}})] (see text).}
\label{fig:norm}
\end{figure}

The normalized experimental $\gamma$-ray strength functions $f=BF/E_{\gamma}^3$
for $^{161,162}$Dy and $^{171,172}$Yb are presented in Fig.\ \ref{fig:norm}. 
The experimental values of the average total radiative width 
$\langle\Gamma_{\gamma}\rangle$ \cite{Mugh} used to determine the normalization
constant $B$ are listed in Table \ref{tab:param}. The figure shows that each 
experimental curve consists of two components. The first one is a smooth 
function of the $\gamma$-ray energy and the second one is connected to a local 
enhancement of the $\gamma$-ray strength function at low $\gamma$-ray energies 
($\sim$3~MeV). The latter component is due to the pygmy resonance, which was 
first observed in (n,$\gamma)$ reactions \cite{Igas1}.

Theoretical curves calculated with the models of Sect.\ III are shown as dashed
curves in Fig.\ \ref{fig:norm}. The parameters adopted in the description of
the GEDR and GMDR are presented in Table \ref{tab:param}. The GEDR parameters
have been determined from the interpolation of systematics over neighboring
isotopes \cite{DB88,GL81}. For the GMDR parameters, there are no rich 
experimental systematics available. Our parameters have been taken from Ref.\ 
\cite{Kope2}, namely $E_{\mathrm{M1}}=41A^{-1/3}$~MeV and 
$\Gamma_{\mathrm{M1}}=4$~MeV\@. The value of $\sigma_{\mathrm{M1}}$ has been 
derived from $f_{\mathrm{E1}}/f_{\mathrm{M1}}$ systematics at $\gamma$-ray 
energies close to the neutron binding energy \cite{Kope3}.

Figure \ref{fig:norm} also shows that the Lorentzian E1 model [label a), Eq.\ 
(\ref{eq:lorentz})] gives an acceptable description for $\gamma$-ray energies 
near $B_n$, in accordance with the systematics of Kopecky and Uhl for deformed 
nuclei \cite{Kope3}. But for lower energies this GEDR model overestimates the 
experimental data. The combination of the KMF E1 model and the Weisskopf M1 
model [label b)] fails to describe the data due to the strong M1 component 
especially in the region of low $\gamma$-ray energies. The KMF model plus the 
Lorentzian M1 model [label c), Eqs.\ (\ref{eq:Kad}, \ref{eq:M1})] is seen to 
give the best description of the general slope of the $\gamma$-ray strength 
function. However, since the M1 strength-function model is generally only
$\sim 20$\% of the GEDR model for the investigated nuclei \cite{Kope3}, no 
further conclusion concerning M1 models could be drawn in this work. In detail,
the agreement of the last curve with the data for $^{161,162}$Dy is 
satisfactory, excluding the low energy region where the pygmy resonance is 
observed. For $^{171,172}$Yb, the slopes of the calculated curves differ 
somewhat from the experimental ones.

In oder to obtain a good parameterization of the $\gamma$-ray strength function
which can fit the experimental data, the sum of the KMF E1 and the Lorentzian 
M1 models has been selected for further modification. In contrast to the common
use, where the nuclear temperature is defined as $T=\sqrt{U/a}$, we will keep 
the temperature fixed (as first proposed by Grudzevich \cite{Gr99,Gr00}) 
according to a constant temperature model of the nuclear level density, which 
is supported by recent findings \cite{Schi2,GH00}. The constant temperature 
model may also be regarded as to mimic the generalized superfluid model of the 
nuclear level density \cite{II79,IW93} in this excitation energy region. The 
mean value of the temperature in the excitation energy region under study is 
$T\sim 0.5$~MeV for the final levels as has been mentioned above. We should 
point out that the temperature dependence of the GEDR width used in the E1 
model is a much disputed topic. Experimental data on damping of the GEDR at low
temperatures ($T<1$~MeV) are absent. At higher temperatures, the damping of the
GEDR is intensively studied with inelastic scattering of light particles (e.g.\
$\alpha$ particles \cite{RA96}) but different theoretical approaches give 
ambiguous results. For example, in the adiabatic coupling model 
\cite{OB96,OB97} the increasing width is explained in terms of thermally 
induced shape fluctuations, yielding in general a $\Gamma\propto\sqrt{T}$ 
dependence. These shape fluctuations become important for $T>1-2$~MeV\@. In the
collisional damping model \cite{YG00} the width of the GEDR is due to 
collisional damping of nucleons, giving a $\Gamma\propto T^2$ law. Experiments 
on $^{208}$Pb show that the data set can be fitted by both of these 
parameterizations \cite{RA96}, while new calculations on the collisional 
damping model using realistic in-medium cross-sections \cite{YG00} show that 
the width is in general underestimated within this model. Also, a recent 
calculation \cite{DE00} on $^{120}$Sn within the phonon damping model shows 
good agreement with experiment.

At temperatures appropriate for the present study ($T\sim 0.5$~MeV), pairing
correlations \cite{DE00} and shell effects \cite{OB96,OB97} have to be taken
into account. Most experimental data on the strength function at the 
low-energetic tale of the GEDR are obtained from ($n,\gamma$) reactions, where
the quadratic temperature-dependence of the GEDR width \cite{Kadm1} is a
popular parameterization \cite{Kope2,Ko93,BC95}. We therefore use this 
parameterization, knowing that the model behind can not account properly for 
the damping mechanism of the GEDR \cite{YG00}.

In order to obtain a good fit of the chosen $\gamma$-ray strength-function 
models to the data we use the temperature as a free parameter because of the 
uncertain temperature dependence of the GEDR width in our temperature region. 
Also, a common normalization constant $K$ was introduced as a free parameter.
Additionally, in order to fit the experimental data in the low-energy region,
a Lorentzian \begin{equation}
f_{\mathrm{py}}(E_\gamma)=\frac{1}{3\pi^2\hbar^2c^2}\frac{\sigma_{\mathrm{py}}
E_\gamma\Gamma_{\mathrm{py}}^2}{(E_\gamma^2-E_{\mathrm{py}}^2)^2+E_\gamma^2
\Gamma_{\mathrm{py}}^2} \label{eq:py}
\end{equation}
with three free parameters $\sigma_{\mathrm{py}}$, $\Gamma_{\mathrm{py}}$ and
$E_{\mathrm{py}}$ has been used for describing the pygmy resonance. Hence, the
fitting function consists of the sum
\begin{equation}
f=K(f_{{\mathrm{E}}1}+f_{{\mathrm{M}}1})+f_{\mathrm{py}},
\label{eq:fitfunc}
\end{equation}
where $f_{{\mathrm{E}}1}$ and $f_{{\mathrm{M}}1}$ are given by Eqs.\ 
(\ref{eq:Kad}) and (\ref{eq:M1}) with parameters from systematics (see Table 
\ref{tab:param}). The values of the fitting parameters for the pygmy resonance,
the normalization constant $K$ and the temperature $T$ are quoted in Table
\ref{tab:fit}. The resulting curves are shown as solid lines in Fig.\ 
\ref{fig:norm}. The agreement with experiment is excellent. It can of course be
debated that other models for the $\gamma$-ray strength functions can be fitted
to the experimental data by letting a sufficient number of parameters free for 
fitting. We have chosen the sum of the KMF E1 and the Lorentzian M1 models with
one fit parameter ($T$) and one free overall normalization constant ($K$) in 
order to obtain a simple parameterization of the experimental data for further 
applications. In addition this parameterization is convenient to extract the 
parameters of the pygmy resonance experimentally.

To judge the relevance of the adopted model of Eq.\ (\ref{eq:fitfunc}), it is 
important to discuss the fitting values obtained. Table \ref{tab:fit} shows 
that the coefficient $K$ is close to 1.0 for all four nuclei\footnote{Here, the
phrase "close to" should be appropriate, since the normalization in other works
is often a factor of 2 uncertain.}. This means that the adopted model 
reproduces the absolute values of the $\gamma$-ray strength functions. The 
small deviation of $K$ from 1.0 can be explained e.g.\ by uncertain values of 
the GEDR parameters for the investigated nuclei, or they can be ascribed 
uncertainties in the experimental normalization of the $\gamma$-ray strength 
functions.

A close inspection of the fitting of Eq.\ (\ref{eq:fitfunc}) to the data shows
that the parameter $T$ is mainly determined by the experimental value of the
$\gamma$-ray strength function in a small $\gamma$-energy region around 
1.5~MeV\@. The fitted values for the nuclear temperature are found around 
$T=0.3$ MeV, a value which is consistent for all four nuclei within the fitting
errors. However, the value is lower than expected from other studies 
\cite{Melb1,Schi2}, giving $T\sim 0.5$~MeV\@. Since the temperature dependence 
of the GEDR is only roughly determined by theory, we therefore interpret $T$ as
a fit parameter not necessarily equal to the nuclear temperature.

\begin{figure}\centering
\includegraphics[totalheight=17.9cm]{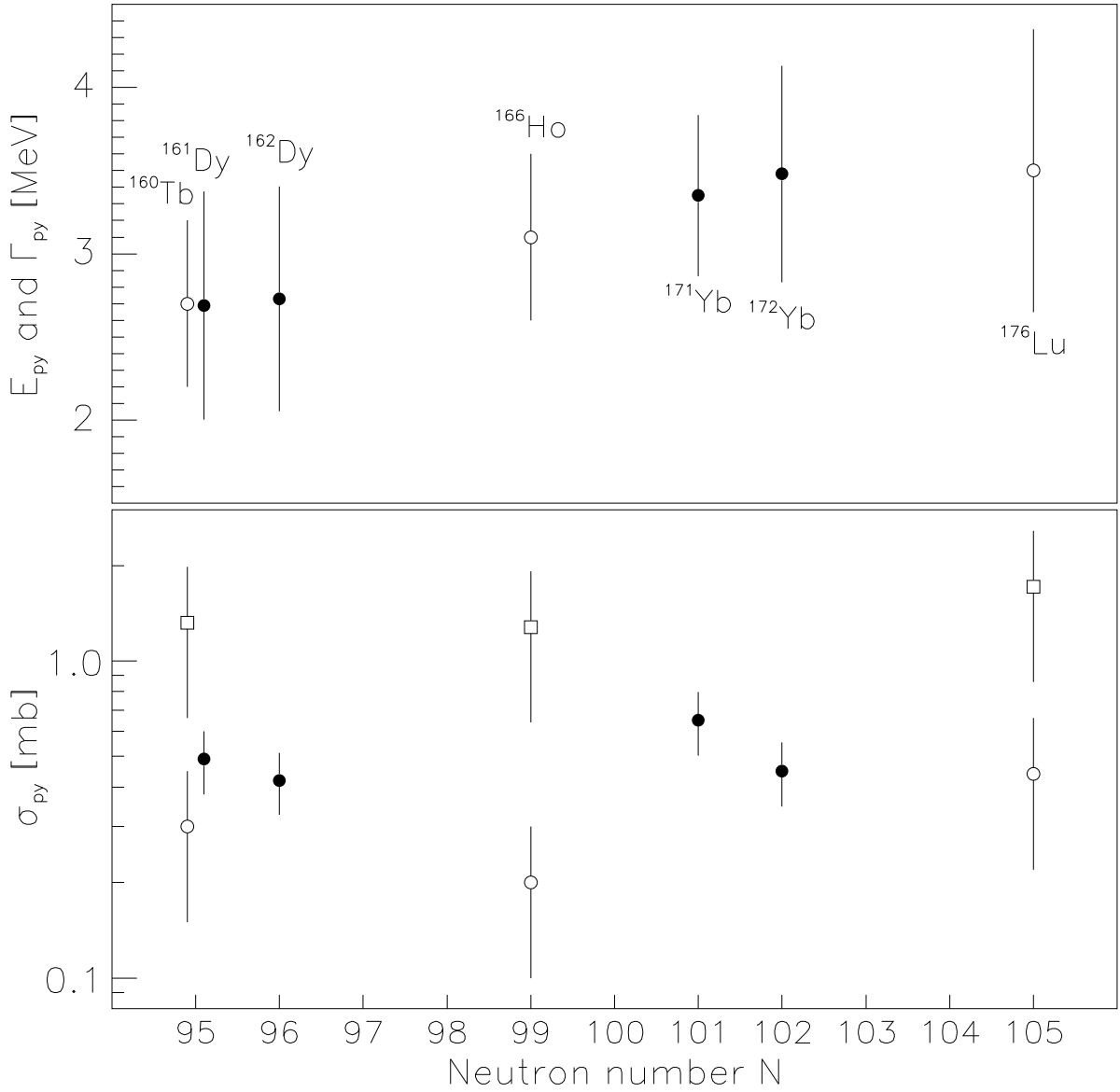}
\caption{Pygmy resonance parameters from the present ($^3$He,$\alpha\gamma$)
reaction (filled circles) compared to those from the (n,$\gamma$) reaction
[8] 
(open circles) as function of neutron number $N$. In the upper panel, the 
resonance energy $E_{\mathrm{py}}$ is displayed as data points and the width 
$\Gamma_{\mathrm{py}}$ is given by the length of the lines through the data 
points. The cross-sections with error bars are shown in the lower panel. For 
the ($^3$He,$\alpha\gamma$) reaction, the quantity $\sigma_{\mathrm{py}}$ is
plotted and assigned an additional systematic error of 20\% from the 
normalization in Eq.\ ({\protect{\ref{eq:norm}}}). For the (n,$\gamma$) 
reaction, the quantities $\sigma_{\mathrm{py}}$ (open circles) and 
$k\sigma_{\mathrm{py}}$ (open squares)
[8] 
are plotted.}
\label{fig:compar}
\end{figure}

It is interesting to compare the pygmy resonance parameters to those obtained
from (n,$\gamma)$ experiments. Unfortunately, the available systematics on
pygmy resonances is very scarce \cite{Igas1} and it is difficult to make
definite conclusions. Nevertheless, Fig.\ \ref{fig:compar} shows that our
fitted pygmy resonance parameters are in good agreement with Ref.\ 
\cite{Igas1}. Both the resonance energy and the width of the pygmy resonances
fit into the systematics. Our strength parameter $\sigma_{\mathrm{py}}$ is
in between the values $\sigma_{\mathrm{py}}$ and $k\sigma_{\mathrm{py}}$ found
in the $({\mathrm{n}},\gamma)$ experiment \cite{Igas1}, see Fig.\ 
\ref{fig:compar}. The comparison supports that the pygmy resonance observed in 
the two reactions has the same physical origin and manifests itself as a common
property of the $\gamma$-ray strength function.

\begin{figure}\centering
\includegraphics[totalheight=17.9cm]{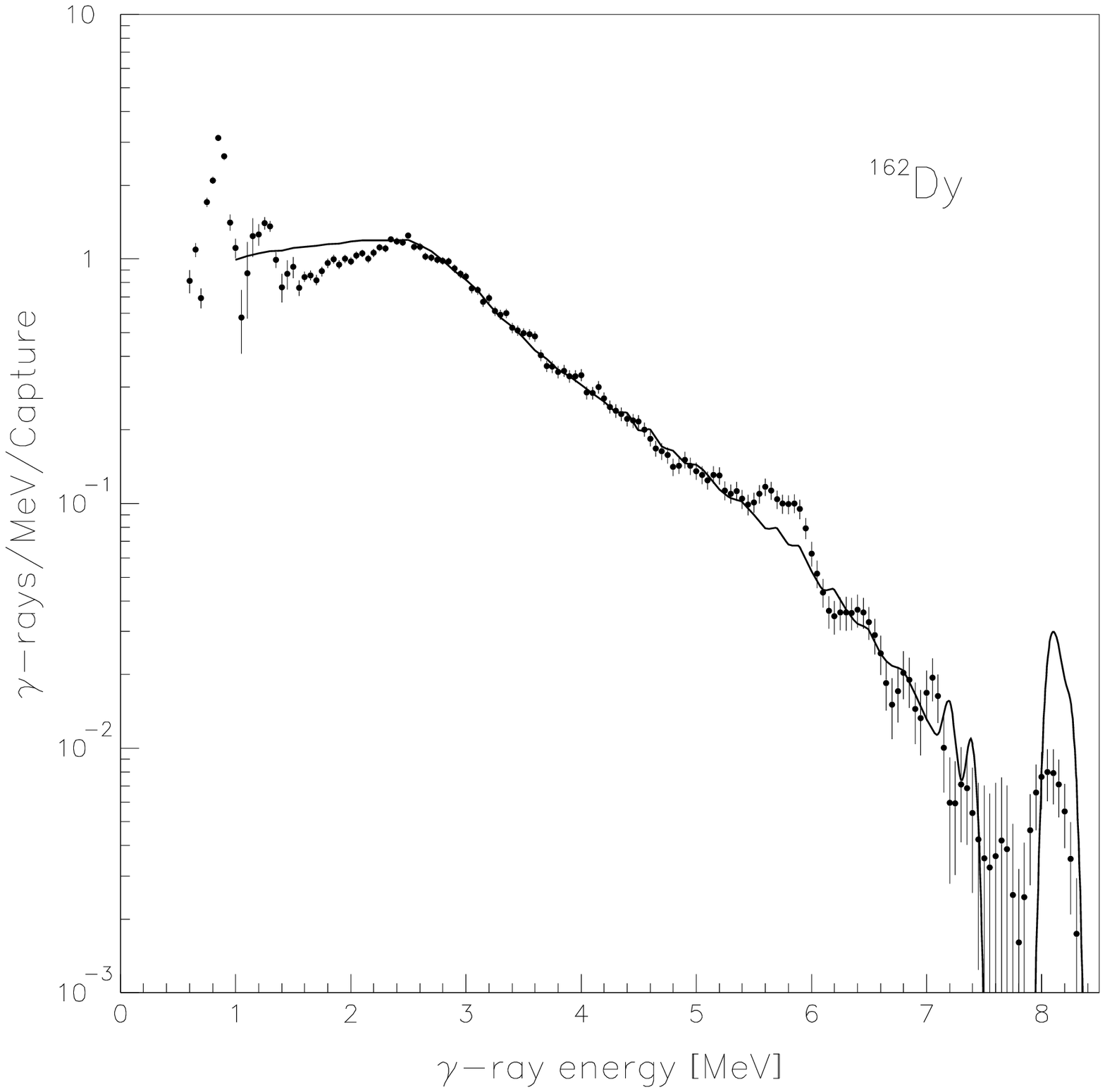}
\caption{The total $\gamma$-ray spectrum for $^{162}$Dy. The data points with
error bars are taken from the $^{161}$Dy(n,$\gamma)^{162}$Dy reaction
[39]
$\gamma$-ray strength function and the level density extracted from the present
$^{163}$Dy$(^3$He,$\alpha\gamma)^{162}$Dy data. The calculation is performed by
averaging over 100~keV intervals.}
\label{fig:igashira}
\end{figure}

As a last test, we have calculated a $\gamma$-ray spectrum using the extracted
$\gamma$-ray strength function $f$ and level density $\rho$ for $^{162}$Dy. The
spectrum includes $\gamma$-ray cascades from an excitation energy of 
$E=B_n+E_n$ and down to the ground state. The spectrum is calculated using a
Monte Carlo simulation, where $f$ and $\rho$ determine the decay pattern. In
Fig.\ \ref{fig:igashira} this spectrum is compared to the $\gamma$-ray spectrum
from the $^{161}$Dy(n,$\gamma)^{162}$Dy reaction (data points with error bars) 
measured at the neutron energy $E_n=47$~keV \cite{Igash162}. Since the 
($^3$He,$\alpha\gamma$) data are based on a broader spin window, details in our
spectra are expected to be more smeared out at final excitation energies having
low level density. This concerns the fine structures seen below 2~MeV and above
5~MeV of $\gamma$-ray energy. The overall agreement between the two spectra is 
gratifying and supports our results.

\section{Conclusions}

The $\gamma$-ray energy dependent factor and the level density for 
$^{161,162}$Dy and $^{171,172}$Yb have been measured using the 
($^3$He,$\alpha\gamma$) reaction. For the first time, the normalized 
$\gamma$-ray strength function $f(E_{\gamma})$ could be extracted from such
data.

Various models are tested against the observed $\gamma$-ray strength function
and the best description is found for the E1 model of KMF with a fixed
temperature plus Lorentzian models for the GMDR and the pygmy resonance. The 
pygmy resonance parameters for $^{161,162}$Dy and $^{171,172}$Yb fit into the
available systematics obtained from (n,$\gamma)$ experiments. Hence, the
adopted approach gives consistent $\gamma$-ray strength functions for the
investigated nuclei.

A few tentative explanations exist for the pygmy resonance. Still, the question
remains open whether the pygmy resonance is of E1 or M1 character. Measurements
of the electromagnetic character of the pygmy resonance is therefore important 
in order to pin down the true nature of this peculiar phenomenon.

\acknowledgements

Stimulating discussions with P.F.~Bortignon are greatly appreciated. We also 
thank M.~Igashira and S.~Mizuno for sending us their data on $^{162}$Dy. The 
authors are grateful to E.A.~Olsen and J.~Wikne for providing the excellent 
experimental conditions. This work is supported by the Norwegian Research 
Council (NFR).

\end{multicols}

\newpage

\begin{table}
\caption{The parameters used for calculation of $\gamma$-ray strength
functions.}
\begin{tabular}{ccccccccccc}
Nucleus&$E^{(1)}_{{\mathrm{E}}1}$&$\sigma^{(1)}_{{\mathrm{E}}1}$&
$\Gamma^{(1)}_{{\mathrm{E}}1}$&$E^{(2)}_{{\mathrm{E}}1}$&
$\sigma^{(2)}_{{\mathrm{E}}1}$&$\Gamma^{(2)}_{{\mathrm{E}}1}$&
$E_{{\mathrm{M}}1}$&$\sigma_{{\mathrm{M}}1}$&$\Gamma_{{\mathrm{M}}1}$&
$\langle\Gamma_\gamma\rangle$\\
&(MeV)&(mb)&(MeV)&(MeV)&(mb)&(MeV)&(MeV)&(mb)&(MeV)&(meV)\\\hline
\rule{0mm}{12pt}$^{161}$Dy&12.13&210&2.6&15.8&250&5.05&7.66&1.60&4.0&108\\
$^{162}$Dy&12.13&210&2.6&15.8&250&5.05&7.65&1.49&4.0&113\\
$^{171}$Yb&12.25&239&2.6&15.5&302&4.80&7.50&1.50&4.0&63\\
$^{172}$Yb&12.25&239&2.6&15.5&302&4.80&7.50&1.76&4.0&75\\
\end{tabular}
\label{tab:param}
\end{table}

\begin{table}
\caption{The parameters obtained from the fit.}
\begin{tabular}{cccccc}
Nucleus&$E_{\mathrm{py}}$&$\sigma_{\mathrm{py}}$&$\Gamma_{\mathrm{py}}$&$T$&
$K$\\
&(MeV)&(mb)&(MeV)&(MeV)&\\\hline
\rule{0mm}{12pt}$^{161}$Dy&2.69(4)&0.49(5)&1.37(22)&0.29(11)&1.34(11)\\
$^{162}$Dy&2.73(5)&0.42(4)&1.35(25)&0.34(10)&1.08(8)\\
$^{171}$Yb&3.35(6)&0.65(7)&0.97(16)&0.34(3) &1.22(10)\\
$^{172}$Yb&3.48(7)&0.45(5)&1.30(23)&0.32(2) &1.24(6)\\
\end{tabular}
\label{tab:fit}
\end{table}

\begin{references}
\bibitem{Blatt}J.M. Blatt and V.F. Weisskopf, \it Theoretical Nuclear Physics
\rm (John Wiley \& Sons, New York, 1952).
\bibitem{DB88}S.S. Dietrich and B.L. Berman, At.\ Data Nucl.\ Data Tables \bf
38\rm, 199 (1988).
\bibitem{Popov}Yu.\ P. Popov, in Neutron Induced Reactions, Proceedings of the
Europhysics Topical Conference, June 21-25, 1982 Smolenice, Institute of
Physics EPRC, Slovak Academy of Sciences, Bratislava 121 (1982).
\bibitem{Kadm1}S.G. Kadmenski{\u{\i}}, V.P. Markushev, and V.I. Furman, Yad.\
Fiz.\ \bf 37\rm, 277 (1983) [Sov.\ J. Nucl.\ Phys.\ \bf 37\rm, 165 (1983)].
\bibitem{Kope2}J. Kopecky and M. Uhl, Phys.\ Rev.\ C \bf 41\rm, 1941 (1990).
\bibitem{Bohr}A. Bohr and B.R. Mottelson, \it Nuclear Structure \rm (W.A.
Benjamin, Inc., Reading, Massachusetts, 1975) Vol.\ II, p.\ 636.
\bibitem{Ko93}J. Kopecky, M. Uhl, and R.E. Chrien, Phys.\ Rev.\ C \bf 47\rm,
312 (1993).
\bibitem{Igas1}M. Igashira, H. Kitazawa, M. Shimizu, H. Komano, and N.
Yamamuro, Nucl.\ Phys.\ \bf A457\rm, 301 (1986).
\bibitem{Berg1}I. Bergqvist and N. Starfelt, Nucl.\ Phys.\ \bf 39\rm, 353
(1962).
\bibitem{Bart1}G.A. Bartholomew, I. Bergqvist, E.D. Earle, and A.J.~Ferguson,
Can.\ J. Phys.\ \bf 48\rm, 687 (1970).
\bibitem{Barr1}R.F. Barrett, K.H. Bray, B.J. Allen, and M.J. Kenny, Nucl.\
Phys.\ \bf A278\rm, 204 (1977).
\bibitem{Joly1}S. Joly, D.M. Drake, and L. Nilsson, Phys.\ Rev.\ C \bf 20\rm,
2072 (1979).
\bibitem{Gutt2}M. Guttormsen, J. Rekstad, A. Henriquez, F. Ingebretsen, and
T.F. Thorsteinsen, Phys.\ Rev.\ Lett.\ \bf 52\rm, 102 (1984).
\bibitem{Riht}D. Bohle, A. Richter, W. Steffen, A.E.L. Dieperink, N. Lo Iudice,
F. Palumbo, and O. Scholten, Phys.\ Lett.\ B \bf 137\rm, 27 (1984).
\bibitem{Hend1}L. Henden, L. Bergholt, M. Guttormsen, J. Rekstad, and T.S.
Tveter, Nucl.\ Phys.\ \bf A589\rm, 249 (1995).
\bibitem{Tvet1}T.S. Tveter, L. Bergholt, M. Guttormsen, E. Melby, and
J. Rekstad, Phys.\ Rev.\ Lett.\ \bf 77\rm, 2404 (1996).
\bibitem{Schi1}A. Schiller, L. Bergholt, M. Guttormsen, E. Melby, J. Rekstad,
and S. Siem, Nucl.\ Instrum.\ Methods Phys.\ Res.\ A \bf 447\rm, 498 (2000).
\bibitem{Melb1}E. Melby, L. Bergholt, M. Guttormsen, M. Hjorth-Jensen, F.
Ingebretsen, S. Messelt, J. Rekstad, A. Schiller, S. Siem, and S.W.
{\O}deg{\aa}rd, Phys.\ Rev.\ Lett.\ \bf 83\rm, 3150 (1999).
\bibitem{Schi2}A. Schiller, A. Bjerve, M. Guttormsen, M. Hjorth-Jensen, F.
Ingebretsen, E. Melby, S. Messelt, J. Rekstad, S. Siem, and S.W.
{\O}deg{\aa}rd, nucl-ex/9909011.
\bibitem{Gutt1}M. Guttormsen, A. Bjerve, M. Hjorth-Jensen, E. Melby, J.
Rekstad, A. Schiller, S. Siem, and A. Beli\'{c}, Phys.\ Rev.\ C \bf 62\rm,
024306 (2000).
\bibitem{Gutt3}M. Guttormsen, A. Atac, G. L{\o}vh{\o}iden, S. Messelt, T.
Rams{\o}y, J. Rekstad, T.F. Thorsteinsen, T.S. Tveter, and Z. Zelazny, Phys.\
Scr.\ \bf T32\rm, 54 (1990).
\bibitem{Brin1}D.M. Brink, Ph.D. thesis, Oxford University, 1955.
\bibitem{Axel1}P. Axel, Phys.\ Rev.\ \bf 126\rm, 671 (1962).
\bibitem{GC65}A. Gilbert and A.G.W. Cameron, Can.\ J. Phys.\ \bf 43\rm, 1446
(1965).
\bibitem{Kope3}J. Kopecky and M. Uhl, in \it Proceedings of a Specialists'
Meeting on Measurement, Calculation and Evaluation of Photon Production Data,
Bologna, Italy 1994 \rm (NEA/NSC/DOC(95)1) p.\ 119.
\bibitem{Mugh}S.F. Mughabghab, \it Neutron cross sections \rm (Academic Press,
New York, 1984), Vol.\ 1, part B.
\bibitem{GL81}G.M. Gurevich, L.E. Lazareva, V.M. Mazur, S.Yu.\ Merkulov, G.V.
Solodukhov, and V.A. Tyutin, Nucl.\ Phys.\ \bf A351\rm, 257 (1981).
\bibitem{Gr99}O.T. Grudzevich, Phys.\ Atom.\ Nucl.\ \bf 62\rm, 192 (1999).
\bibitem{Gr00}O.T. Grudzevich, Phys.\ Atom.\ Nucl.\ \bf 63\rm, 414 (2000).
\bibitem{GH00}M. Guttormsen, M. Hjorth-Jensen, E. Melby, J. Rekstad,
A. Schiller, and S. Siem, Phys.\ Rev.\ C \bf 61\rm, 067302 (2000).
\bibitem{II79}A.V. Ignatyuk, K.K. Istekov, and G.N. Smirenkin, Yad.\ Fiz.\ \bf
29\rm, 875 (1979) [Sov.\ J. Nucl.\ Phys.\ \bf 29\rm, 450 (1979)].
\bibitem{IW93}A.V. Ignatyuk, J.L. Weil, S. Raman, and S. Kahane, Phys.\ Rev.\ C
\bf 47\rm, 1504 (1993).
\bibitem{RA96}E. Ramakrishnan \sl et al.\ \rm Phys.\ Lett.\ B \bf 383\rm, 252
(2000).
\bibitem{OB96}W.E. Ormand, P.F. Bortignon, and R.A. Broglia, Phys.\ Rev.\
Lett.\ \bf 77\rm, 607 (1996).
\bibitem{OB97}W.E. Ormand, P.F. Bortignon, R.A. Broglia, and A. Bracco, Nucl.\
Phys.\ \bf A614\rm, 217 (1997).
\bibitem{YG00}Osman Yilmaz, Ahmet Gokalp, Serbulent Yildirim, and Sakir Ayik,
Phys.\ Lett.\ B \bf 472\rm, 258 (2000).
\bibitem{DE00}N. Dinh Dang, K. Eisenman, J. Seitz, and M. Thoennessen, Phys.\
Rec.\ C \bf 61\rm, 027302 (2000).
\bibitem{BC95}F. Be{\v{c}}v{\'{a}}{\v{r}}, P. Cejnar, J. Honz{\'{a}}tko, K.
Kone{\v{c}}n{\'{y}}, I. Tomandl, and R.E. Chrien, Phys.\ Rev.\ C \bf 52\rm,
1278 (1995).
\bibitem{Igash162}Satoshi Mizuno, Masayuki Igashira, and Koji Masuda, J. Nucl.\
Sci.\ Technol.\ \bf 36\rm, 493 (1999).
\end{references}
\end{document}